# Ink-jet printed 2D crystal heterostructures

Francesco Bonaccorso

Istituto Italiano di Tecnologia, Graphene Labs,
Via Morego 30, 16163 Genova, Italy
francesco.bonaccorso@iit.it

*Abstract*— **The availability of graphene and related two-dimensional (2D) crystals in ink form, with on demand controlled lateral size and thickness, represents a boost for the design of printed heterostructures. Here, we provide an overview on the formulation of functional inks and the current development of inkjet printing process enabling the realization of 2D crystal-based heterostructures.**

## I. INTRODUCTION

Heterostructures have already played a central role in technology, for the realization of, *e.g.*, semiconductor lasers, high mobility field effect transistors (FETs) and diodes. Heterostructures based on two-dimensional (2D) crystals (2DHs), *i.e.* isostructural system of 2D crystals assembled by stacking different atomic planes in sandwiched structures, offer the prospect of extending existing technologies to their ultimate limit using monolayer-thick tunnel barriers and quantum wells.[1] In fact, the on-demand assembly of 2D crystals allows the engineering of artificial three-dimensional (3D) crystals, exhibiting tailor-made properties that could be tuned to fit any application. Pioneering research has already demonstrated first proof of principle 2DHs, such as field effect vertical tunneling transistors based on graphene with atomically thin hexagonal boron nitride (*h*-BN) acting as a tunnel barrier,[2] gate-tunable p–n diodes based on a p-type BP/n-type monolayer molybdenum disulphide ($MoS_2$)[3] or proof-of-concept photovoltaic cells.[4]

Although it is not difficult to envision many possible combinations of materials, one stack of many different layers with atomic precision, the practical realization of such vision is much more complicated. The ideal approach would be to directly grow 2DHs where needed, but this target is still far from any practical realization. Currently, three methods have been exploited for the production of 2DHs: (I) layer by layer stacking via mechanical transfer;[5,6] (II) direct growth by chemical vapour deposition (CVD)[7] and molecular beam epitaxy (MBE)[8]; and (III) layer by layer deposition of solution processed 2D crystals. However, at this time all of the aforementioned approaches have limitations. The layer by layer stacking or deterministic placement (I) via mechanical transfer relies on the mechanical exfoliation of layered materials into atomically thin sheets.[9] Moreover, in order to fabricate 2DHs with clean interfaces (*i.e.*, without trapped adsorbates between the stacked layers), which is necessary for long-term device reliability, a dry transfer procedure is preferred. This would avoid the wet conditions with polymer coatings, which suffer from polymer contamination. Even the most developed dry transfer protocols may not result in perfectly clean interfaces, as some adsorbates may get trapped between the stacked 2D layers.[10] Although this procedure is now optimized to yield sophisticated layered structures, it is limited to vertical structures,[11] it is not suitable for layer registration with the underlying films and, even more critical, it is impractical for high volume manufacturing. The direct growth (II) of different 2D crystals vertically stacked is another approach suitable for the production of 2DHs. First attempts have already shown the feasibility of such processes. Just to highlight some examples, *h*-BN has already been demonstrated to be an effective substrate for the CVD growth of graphene.[12] Vertically stacked 2DHs have been synthesized by the sequential CVD growth of 2D transition metal dicalchogenides (TMDs) on top of pre-existing *h*-BN[13] and graphene[14,15] or by the selenization and sulfurization of elemental metals.[16,17] The co-reaction of Mo and W-containing precursors with chalcogens[18] or the *in-situ* vapor-solid reactions[19] have proven to be other feasible routes for the realization of lateral and vertical 2DHs. Van der Waals epitaxy has also been exploited, using $WCl_6$/S, $MoCl_5$/S and Se as precursors and $SnS_2$ templates.[20] However, these approaches have significant limitations in that monolayer by monolayer growth process conditions have not been established yet. That is, island or 3D growth is observed rather than 2D growth in contrast to what has been observed in CVD graphene growth.[21] Any industrial application will require a scalable approach. To this aim, layer-by-layer deposition (III) from 2D crystal-based inks (Fig. 1)[22,23] could be the right strategy for scalable production of 2DHs.

Here we will present the latest progress on the large-scale placement of 2D crystal-based inks by inkjet printing,[24] which allows printing of layers of different 2DHs on a large scale. We will discuss several issues that need to be optimized, such as the uniformity of large area film stacks, the discontinuity of the individual crystals assembling the 2DHs, which are currently affecting the 2DH (opto)electronic properties.

## II. INKJET PRINTING OF 2D CRYSTALS

Inkjet printing is used to print a wide range of (opto)electronic devices.[25,26,27,28,29,30] Many factors influence the printed features. In fact, during an inkjet printing process, a regular jetting from the print-head nozzle is needed to avoid printing instabilities, *i.e.*, formation of satellite drops and jetting deflection.[31] Depending on the ink wettability behaviour at the nozzle, unwanted spray formation may occur instead of a regular jetting.[32] Furthermore, the resolution of the printed feature is influenced by the drop velocity $v$ (*i.e.* 5-10 ms$^{-1}$) when it impacts onto the substrate.[32]

The formulation of 2D crystals-based (as well as nanomaterials in general) printable inks is rather challenging because the various liquid properties such as density ($\rho$), surface tension ($\gamma$) and viscosity ($\eta$) have a strong effect on the printing process itself.[33] These ink physical properties need to be carefully tuned and can be summarized in dimensionless figures of merit (FoM) such as: the Reynolds ($N_{Re}$) and Weber ($N_{We}$) numbers,[34,35] and the inverse of the Ohnesorge number, $Z$ ($1/N_{Oh}$), defined as the ratio between the $N_{Re}$ and the square root of the $N_{We}$.[33] Different $Z$ values for the stable drop formation have been proposed,[36,37] with $Z$ values mostly enclosed in the range $4 \leq Z \leq 14$, although many reports have also demonstrated stable ink-jet printing with $Z$ values of the printing ink outside this range.[38,39,40,41] In particular, nanomaterial-based inks (e.g., polystyrene nanoparticle[38] and graphene-based inks[42]) have also been ink-jet printed with $Z$ values outside the aforementioned range. The morphological properties of the nanomaterials (e.g., lateral size for 2D crystals) dispersed in the ink as well as the formation of aggregates in the ink and their accumulation on the print-head can also contribute to printing instabilities. Dispersed nanomaterials with lateral sizes smaller than ~1/50 of the nozzle diameter (*i.e.*, $\geq 100\mu m$[43]) can reduce these damaging effects.[44,45] Moreover, wetting and adhesion[31] to the substrate and its distance to the nozzle (*e.g.*, 1-3 mm)[46] are other key requirements for the printing. The behaviour of a droplet which spreads on the substrate under the action of the inertia and surface forces is characterized by the dynamic contact angle $\theta_c$,[47] a parameter linked with the substrate wettability.

Graphene and related 2D crystals are emerging as promising functional materials for ink formulation.[32,42] The first attempts in formulating 2D crystal-based inks for inkjet printing exploited graphene oxide (GO) or reduced graphene oxide (RGO).[48,49,50,51,52,53,54,55,56,57,58,59] Although several processes have been developed to chemically "reduce" the GO flakes in order to re-establish an electrical and thermal conductivity as close as possible to pristine graphene, RGO contains structural defects.[60] Liquid phase exfoliation (LPE) of pristine graphite[61] to obtain un-functionalized graphene flakes is a most promising approach for the formulation of graphene-based inks.[32] Most importantly, LPE allows the formulation of other 2D-crystal-based inks, see Figure 1, starting from the exfoliation of their bulk counterpart.[62]

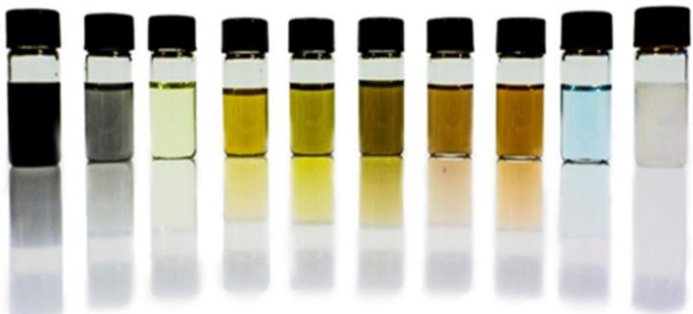

**Figure 1**: 2D-crystal-based functional inks.

In this case, the process is mostly driven by the choice of solvents able to disperse the flakes.[31,34] The first formulation involved the use of graphene inks prepared in N-Methyl-2-pyrrolidone (NMP)[42] and Dimethylformamide (DMF)[63] to print conductive stripes, achieving sheet resistance ($R_S$) values of ~30 k$\Omega$/☐ on glass. The formation of coffee ring effect when graphene ink is printed on rigid substrates (glass[64] and SiO$_2$[62]) can be overcome by substrate treatments, e.g., hexamethyldisilazane. Alternative routes to avoid coffee ring effects can be either the use of low boiling point solvents, with higher enthalpy of vaporization than water, or substrates that promote adhesion.[65] In the first case, the use of low boiling point solvents for the exfoliation of layered crystals has to take into account the mismatch between the $\gamma$ of the solvent and the surface energy of the sheets. This issue could be overcome by the exploitation of co-solvents,[66,67] *e.g.*, water/alcohol mixtures, to tune the fluidic properties of the liquid. This allowed for the direct inkjet printing of graphene-based conductive stripes from low boiling point solvents also on flexible substrates achieving $R_S$~1-2 K$\Omega$/☐.[32,68]

The inkjet printing technology has been also recently demonstrated a promising tool to print other 2D crystals (e.g., MoS$_2$, WS$_2$) apart graphene, overcoming several still existing drawbacks for a reliable mass production of high-quality 2D crystal-based films/patterns,[69,70,71] see Figure 2. Despite these progresses, several issues need to be still overcome for the optimization of 2D crystal-based ink-jet printing. The main problem is that the common solvents used in LPE (e.g., DMF and NMP) are toxic and have very low viscosities ($< 2$ mPa·s), the latter strongly decreasing the jetting performance. In addition, the concentration of 2D crystals in these solvents is low ($< 1$ g L$^{-1}$), thus requiring many printing passes to obtain functional films. Another issue to be faced, especially with the use of high boiling point solvents is the required post-processing annealing for solvent removal,[72] which poses severe limitations to the type of substrate to be used for the printing process. Similar issues are also faced in the case of 2D crystals-based inks prepared in aqueous solution, where the surfactants/polymers removal requires thermal and/or chemical treatments,[73] which are often not compatible with the substrate.

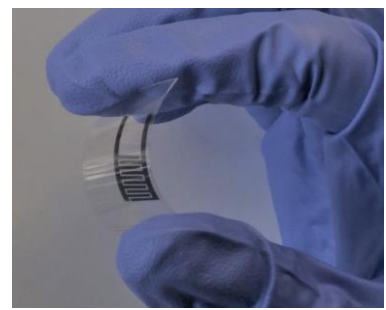

**Figure 2**: Inkjet-printed interdigitated graphene supercapacitor on PET (100 printing layers).

III. PRINTED HETEROSTRUCTURES

The practical realization of 2DHs, to obtain stacks of many different layers with atomic precision, is a difficult task especially in view of industrial application, which requires a scalable approach. In this context, "layer-by-layer deposition" from 2D crystal-based inks by means of Langmuir–Blodgett,[24] and inkjet printing[69] allows the deposition of layers of different 2D crystal-based heterostructures on a large scale. Although the printing approach is very recent, some proofs of concept devices have already been demonstrated.[33] In this context, the simplest structure is an all-inkjet printed in-plane photodetector based on $MoS_2$ channel and interdigitated graphene electrodes.[69] Another example is a tunneling transistor, where the tunneling between the top and bottom graphene layer through a TMD layer is back gate controlled.[74]

A key advantage of this approach could be represented by the possibility to integrate/complement other production approaches, for example for the realization of contacts. Very recently, a programmable logic memory device (*i.e.*, graphene/$WS_2$/graphene) has been realized by inkjet printing technology.[75] An all-printed, vertically stacked transistors with graphene source, drain, and gate electrodes, a TMD channel, and a BN separator has been demonstrated.[78] The proposed printed vertical heterostructure has shown a charge carrier mobility of 0.22 $cm^2/Vs$,[76] which is however rather low. Thus new insights into the assembly of such printed heterostructures are needed to further improve the performances of such devices.

IV. CONCLUSION

The realization of printed heterostructures based on 2D crystal-inks is now emerging as a possible route for their large scale production. However, this technology is still in its infancy and several issues have to be solved. In fact, apart from the uniformity of large area film stacks, the assembly of such heterostructures suffers from discontinuity of the individual crystals, thus resulting in structures with (opto)electronic properties of lower performance with respect to the one obtained by dry transfer methods such as layer by layer stacking *via* mechanical transfer[5] and direct growth.[7,8] Thus, before the layer-by-layer deposition of 2D crystal-based dispersions and inks can be exploited for the realization of vertical 2D heterostructures with (opto)electronic properties comparable with the ones achieved with the other approaches, a strong experimental effort is needed to fully evaluate the potentiality of this method, overcoming the aforementioned issues.

Another issue to tackle, as in the case of layer-by-layer stacking via mechanical transfer, relies on the fact that the layer-by-layer deposition of 2D crystal-based dispersions and inks can be exploited for the realization of vertical 2DHs but not for lateral ones, which is a main limitation for this approach.

ACKNOWLEDGMENT

We thank A. Ansaldo, N. Curreli, A. E. Del Rio Castillo, E. Petroni for useful discussion. This work was supported by the European Union's Horizon 2020 research and innovation program under grant agreement No. 696656—GrapheneCore1.